\documentclass[]{emulateapj}

\makeatletter

\newenvironment{figurehere}
  {\def\@captype{figure}}
  {}
\makeatother

\def\wcen{$\omega$~Cen}

\def\x{$\times$}
\def\about{$\sim$}

\def\simgt{\buildrel{>}\over \sim}
\def\simless{\buildrel{<}\over \sim}
\def\simgreat{\buildrel{>}\over \sim}

\def\simgt{$\ga$}

\def\spt{$\buildrel{\rm s}\over .$}
\def\asec{$^{\prime\prime}$}
\def\amin{$^\prime$}
\def\secspt{$\buildrel{\prime\prime}\over .$}
\def\spt{$\buildrel{\prime\prime}\over .$}
\def\minspt{$\buildrel{\prime}\over .$}

\def\mass{${\cal M}$}
\def\msun{${\cal M}_{\odot}$}
\def\msunyr{${\cal M}_{\odot}$ yr$^{-1}$}
\def\sun{$_{\odot}$}
\def\yr-1{yr$^{-1}$}
\def\mb435{$B_{435}$}
\def\mr625{$R_{625}$}
\def\MB435{$M_{435}$}
\def\MR625{$M_{625}$}
\def\br{$B_{435}-R_{625}$}
\def\intbr{($B_{435}-R_{625})_0$}
\def\hr{$H\alpha-R_{625}$}
\def\ha{H$\alpha$}

\def\subr #1{_{{\rm #1}}}

\def\fxfv{f$_{\rm X}$/f$_{\rm V}$}

\slugcomment{Accepted for publication in {\it The Astrophysical Journal}.}

\shorttitle{qLMXB in \wcen}
\shortauthors{Haggard et al.}

\begin{document}

\title{HST/ACS Imaging of Omega Centauri: Optical Counterpart
for the Quiescent Low-Mass X-Ray Binary\altaffilmark{1}}

\altaffiltext{1}{Based on observations with the NASA/ESA Hubble Space
                 Telescope obtained at the Space Telescope Science 
                 Institute, which is operated by the Association of
                 Universities for Research in Astronomy, Incorporated,
                 under NASA contract NAS 5-26555}

\author{Daryl Haggard\altaffilmark{2,3}, 
        Adrienne M. Cool\altaffilmark{2},
        Jay Anderson\altaffilmark{4}, 
        Peter D. Edmonds\altaffilmark{5},
        Paul J. Callanan\altaffilmark{6}, 
        Craig O. Heinke\altaffilmark{5}, 
        Jonathan E. Grindlay\altaffilmark{5}, and
        Charles D. Bailyn\altaffilmark{7}}

\altaffiltext{2}{Dept. of Physics and Astronomy, San Francisco State 
University, 1600 Holloway Ave., San Francisco, CA 94618; cool@sfsu.edu}
\altaffiltext{3}{Dept. of Astronomy, University of Washington, Box 351580, 
Seattle, WA 98195; dhaggard@astro.washington.edu}
\altaffiltext{4}{Dept. of Physics and Astronomy, Rice University, 6100 
Main St., Houston, TX 77005; jay@eeyore.rice.edu}
\altaffiltext{5}{Harvard-Smithsonian Center for Astrophysics, 60 Garden 
St., Cambridge, MA 02138; pedmonds@cfa.harvard.edu, 
cheinke@cfa.harvard.edu, josh@cfa.harvard.edu}
\altaffiltext{6}{Department of Physics, University College, Cork, Ireland; 
paulc@ucc.ie}
\altaffiltext{7}{Department of Astronomy, Yale University, New Haven, CT 
06520; bailyn@astro.yale.edu}

\begin{abstract}

We report the discovery of an optical counterpart to a quiescent
neutron star in the globular cluster $\omega$ Centauri (NGC~5139).
The star was found as part of our wide-field imaging study of \wcen\
using the Advanced Camera for Surveys (ACS) on Hubble Space Telescope.
Its magnitude and color (\mr625 = 25.2, \br = 1.5) place it more than
1.5 magnitudes to the blue side of the main sequence.  Through an \ha\
filter it is \about 1.3 magnitudes brighter than cluster stars of
comparable \mr625\ magnitude.  The blue color and \ha\ excess suggest
the presence of an accretion disk, implying that the neutron star is
accreting from a binary companion and is thus a quiescent low-mass
X-ray binary.  If the companion is a main-sequence star, then the
faint absolute magnitude (\MR625 $\simeq$ 11.6) constrains it to be of
very low mass (\mass $\simless$ 0.14\msun).  The faintness of the disk
(\MB435 \about 13) suggests a very low rate of accretion onto the
neutron star.  We also detect 13 probable white dwarfs and three
possible BY Draconis stars in the 20\asec\ \x\ 20\asec\ region
analyzed here, suggesting that a large number of white dwarfs and
active binaries will be observable in the full ACS study.

\end{abstract}


\section{Introduction}

Omega Centauri is a prime target for studies of stellar collisions.
It is nearby (D $\simeq$ 5 kpc), massive and, despite a relatively
moderate central density, has one of the highest predicted rates of
stellar interactions among globular clusters, owing to its very large
core.  Channels for compact binary production include exchange
encounters, binary-binary collisions, and possibly tidal capture (Hut
et al.\ 1992, Fregeau et al.\ 2003, and references therein).
Di~Stefano and Rappaport (1994) predicted that \about 100 cataclysmic
variables (CVs) formed by tidal capture should be in \wcen\ at
present.  Some compact binaries may also evolve directly from
primordial binaries in this cluster (Davies 1997).

We are using the Chandra X-Ray Observatory and the Advanced Camera for
Surveys (ACS) on the Hubble Space Telescope (HST) to search for
compact binary stars in \wcen.  The nine fields observed with ACS form
a mosaic that encompasses more than 100 of the Chandra sources
identified with ACIS-I in the direction of \wcen\ (Cool, Haggard, \&
Carlin 2002).  The ACS mosaic contains over a million stars and
represents the most complete census of stars in \wcen\ yet obtained.

Here we report the first detection of an optical counterpart for a
Chandra source using the ACS data.  The source in question was
identified as a possible transient neutron star in quiescence by
Rutledge et al.\ (2002) in a spectral analysis of 32 of the brightest
sources in the Chandra data.  Its X-ray spectrum was found to be
consistent with thermal emission from a neutron star with a hydrogen
atmosphere and inconsistent with several other possible explanations.
The characteristics of the optical counterpart that we have identified
suggest the presence of a disk and binary companion, as in a quiescent
low-mass X-ray binary (qLMXB).  While more than two dozen qLMXBs have
been identified in ten different globular clusters from their X-ray
spectra (Heinke et al.\ 2003b), this is only the
second optical counterpart found for one of these objects during
quiescence (the other being X5 in 47~Tuc; Edmonds et al.\ 2002). qLMXBs
are believed to play an important role in the production of millisecond
pulsars; studying them in globular clusters also holds the promise of
new constraints on neutron star equations of state (Brown, Bildsten \&
Rutledge 1998, Heinke et al.\ 2003a).

We describe the ACS observations and our astrometric and photometric
analyses in \S~2.  In \S~3 we present the proposed optical
counterpart and discuss it in \S~4.  Results of our search for
additional optical counterparts to Chandra sources, as well as our
study of stellar populations in \wcen\ using the ACS data, will appear
in subsequent papers.


\begin{deluxetable*}{ccccccc}[t]
\tabletypesize{\small}
\tablecaption{Astrometry}
\tablewidth{0pt}
\tablecolumns{7}

\tablehead{
       & Chandra R.A. & Chandra Decl. & opt.$-$X-ray, raw & opt.$-$X-ray, 
corr. & HST archival &  pixel coords \\
Source &   (J2000)    &   (J2000)     & $\Delta\alpha$, $\Delta\delta$ & 
$\Delta\alpha^\prime$, $\Delta\delta^\prime$ & image, chip & x, y}

\startdata
 XA    & 13 26 52.141 &  -47 29 35.63 &  $-$0\spt 08,   \phs 0\spt 31 & 
$-$0\spt 03,   \phs 0\spt 02 & j6lp05vsq, 1 & \phn 323, \phn 864 \\
 XB    & 13 26 53.513 &  -47 29 00.37 &  $-$0\spt 02,   \phs 0\spt 28 & 
\phs 0\spt 03, $-$0\spt 02   &      ``      & 1005, 1194  \\
 XC    & 13 26 48.656 &  -47 27 44.88 &  \phs 0\spt 00, \phs 0\spt 33 & 
\phs 0\spt 05, \phs 0\spt 04 &      ``      & 2682, \phn 280 \\
 qLMXB & 13 26 19.795 &  -47 29 10.64 &  \phs 0\spt 19, \phs 0\spt 35 & 
\phs 0\spt 24, $-$0\spt 05   & j6lp02ffq, 2 & 1828, \phn 325 \\ 
\enddata

\end{deluxetable*}

\section{Observations and Analysis}

The ACS observations were made on June 27--29, 2002.  They consist of
a 3\x 3 mosaic of 9 pointings with the Wide Field Camera (WFC)
covering about 10\amin\ \x 10\amin, out to beyond \wcen's half-light
radius (r$\subr{h}$ = 288\asec; Harris 1996).  The X-ray source of
interest here lies \about 4\minspt 5 west of the cluster center.  At
each of the pointings we obtained 3 \x\ 340s exposures with the F625W
(\mr625) and F435W (\mb435) filters and 4 \x\ 440s exposures with the
F658N (\ha) filter, shifting the camera between exposures to fill the
chip gap.  Having 3 to 4 exposures per filter enables us to recognize
and eliminate false detections and poor measurements caused by cosmic
rays.  One short exposure in each of \mr625\ and \mb435\ was also
taken to fill out the cluster's horizontal and giant branches.  We use
these filters to look in the X-ray error circle for stars that are
blue and/or \ha-bright, as potential signatures of accretion.

\subsection{Astrometry}

Chandra coordinates determined using $wavdetect$
(http://asc.harvard.edu/ciao) for the quiescent neutron star
identified by Rutledge et al.\, and for three previously known X-ray
sources in the core, are given in Table 1, columns 2 and 3.  To map
these positions onto the ACS/WFC images, we first applied a distortion
correction to the individual WFC images using the solution obtained
for the $B_{475}$ filter from a study of 47~Tuc (Anderson 2002).  This
solution should be accurate to \about\ 0.15 WFC pixel (1 WFC pixel =
0\secspt 05).  We then stitched together all the individual \mb435\
frames to make a mosaic of the entire field.  The images fit together
well, with no sign of misaligned star images where chips overlap,
indicating that the distortion correction is working well.

To determine the R.A. and Dec. associated with stars on the mosaic, we
used the star lists of Kaluzny et al.\ (1996) and van Leeuwen et al.\
(2000).  More than 15,000 of the former and 4000 of the latter fall
within the field of view of our mosaic.  There is a small zeropoint
offset of 0\secspt 5 between the two systems, which we take to be
indicative of the uncertainty in the absolute frame; we used the
Kaluzny system to define the transformation between our mosaic system
and R.A. and Dec.  We estimate that the transformation is accurate to
$\simless$ 0\secspt 4 over the entire mosaic.

We cross-checked the optical coordinates using counterparts previously
identified in HST/WFPC2 images for two ROSAT sources in the cluster
core, XA and XB.  We recovered stars A and B that Carson, Cool, and
Grindlay (2000) identified as CVs, and confirmed that both are
\ha-bright and blue in the ACS/WFC images (Haggard et al.\ 2002).  The
differences between our optical positions for these two stars and the
Chandra positions (Table 1, column 4) are well within the 0\secspt 6
uncertainty (90\% confidence) in Chandra's absolute coordinate system.

Noting that the X-ray positions for stars A and B are offset from
their optical positions in nearly the same direction, we used them to
to improve the placement of the other X-ray sources on the ACS mosaic.
We shifted the X-ray positions northwest by ($\Delta\alpha$,
$\Delta\delta$) = ($-$0\secspt 05, 0\secspt 30).  This shift places
star C, identified by Carson et al.\ (2000) as a tentative counterpart
of the third ROSAT core source, XC, just $\Delta$r$^\prime$ = 0\secspt
06 from the corresponding Chandra source position (Table 1, column 5),
confirming that it is the likely source of the X-rays.  Star C is
\ha-bright but not blue; it may be either a BY Dra-type active binary
or a cataclysmic variable.

\begin{figurehere}
\vspace*{0.8cm}
\centerline{\resizebox{3in}{!}{\includegraphics{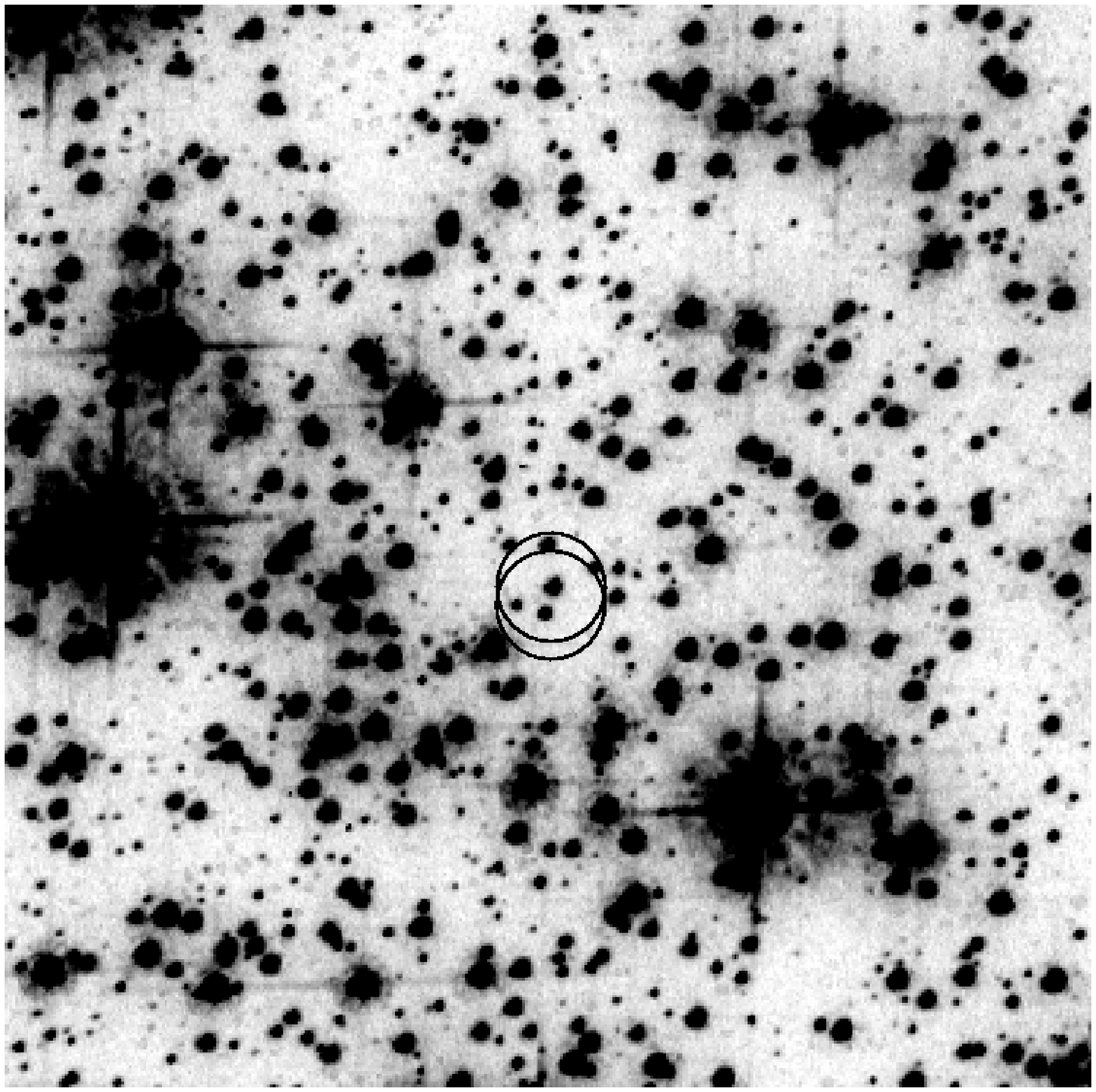}}}
\vspace*{0.2cm}
\caption{The 20\asec\ \x\ 20\asec\ region used in the search for the
optical counterpart to the qLMXB in the \mb435\ filter.  North is up
and east to the left, approximately.  This field is roughly 4\minspt 5
west of the cluster center.  Pre- and post-boresite correction 1\asec\
radius X-ray error circles are shown (lower and upper circle,
respectively).}
\vspace*{0.8cm}
\label{fig01}
\end{figurehere}

The effect of this boresite correction on the position of the X-ray
error circle for the qLMXB is shown in Fig.\ 1.  The remaining
uncertainty in the position of the qLMXB is likely to be dominated by
the uncertainty in off-axis $wavdetect$ positions, \about 0\secspt 5
at the 4\minspt 5 off-axis angle of the object in the Chandra ACIS-I
observation (Feigelson et al.\ 2002).  An additional uncertainty of up
to \about 0\secspt 4 associated with the construction of the
ACS mosaic and the large separation between the qLMXB and stars A and B
may also be present.  We therefore adopted a 1\asec\ error circle in
the search for its optical counterpart.

\subsection{Photometry}

In view of the variability of the point spread function (PSF) in the
ACS/WFC (Instrument Science Report ACS 2003-06), we extracted a 400
\x\ 400 pixel area (\about 20\asec\ \x\ 20\asec) approximately
centered on the Chandra source position from each of the 12 images
(Fig.\ 1).  We used DAOPHOT (Stetson 1987) for the analysis, as the
crowding is significant even outside the cluster core.  We selected
\about 10 bright, isolated, unsaturated stars to characterize the PSF.
As it demonstrated little or no variability across this small region,
we adopted a constant model.  The best results were obtained by making
a separate PSF for each image.

We obtained a preliminary star list using DAOPHOT/FIND and then
examined and compared the images by eye in order to remove cosmic rays
and other spurious detections.  We also manually added to the star
list near neighbors revealed in the course of PSF-fitting.  Magnitudes
and positions for the 1454 stars identified in this way were then
obtained using DAOPHOT/ALLSTAR for each image independently.  Lastly,
we matched up stars measured in each of the 10 long exposures,
requiring that the positions match to within 1 pixel.  We also
analyzed the images using ALLFRAME (Stetson 1994), which is designed
to use consistent positions for stars in all images.  ALLSTAR gave
better results, apparently because of the difficulty in transforming
star positions between non-aligned exposures with sufficient accuracy,
due to the large distortion in the WFC.  Here we report results
obtained using ALLSTAR, which we have calibrated using ``vegamag''
zeropoints kindly provided by Sirianni et al.\ (2003).

\begin{figurehere}
\vspace*{0.5cm}
\centerline{\resizebox{3in}{!}{\includegraphics{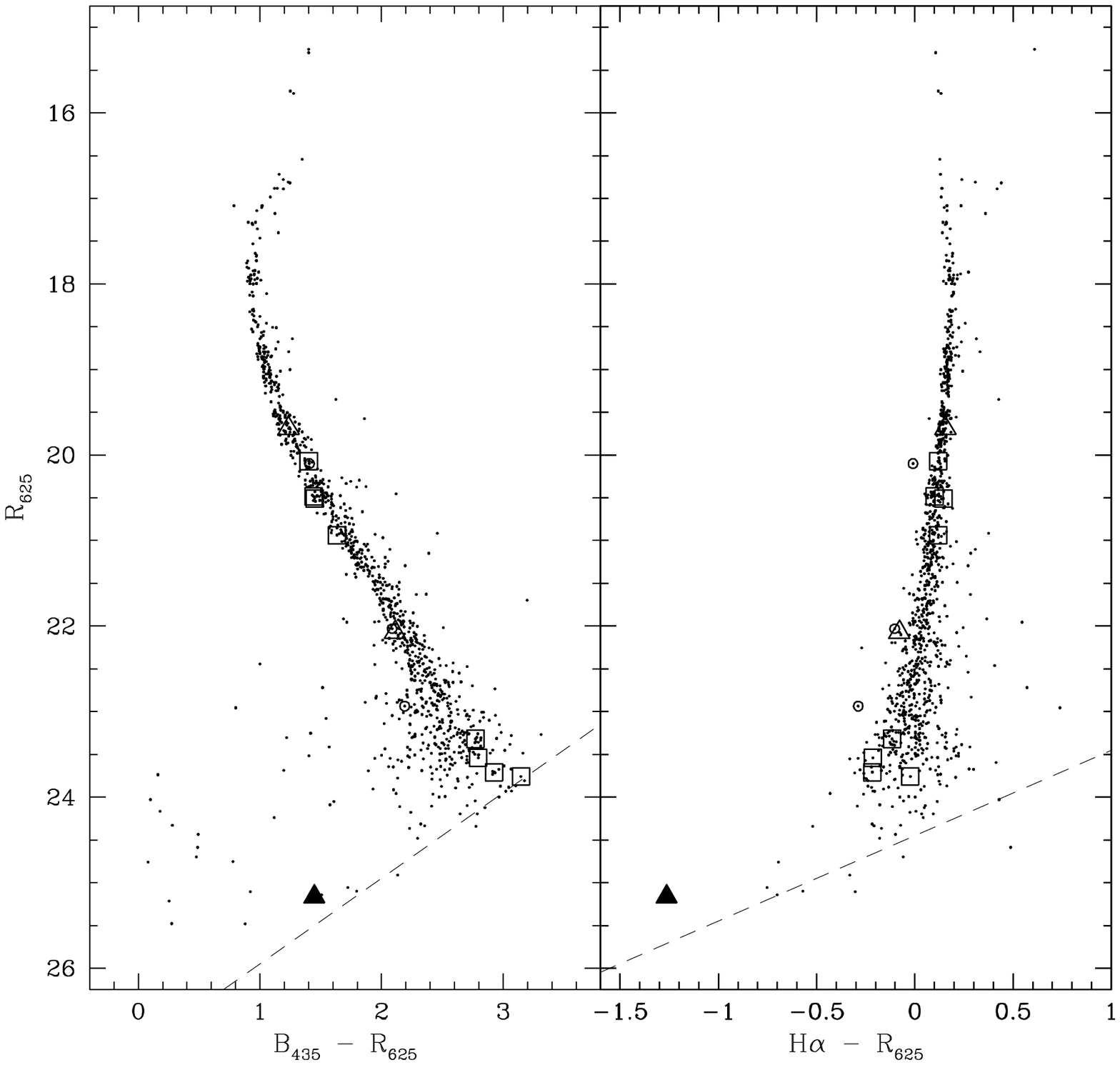}}}
\vspace*{0.2cm}
\caption{CMDs for stars in the region shown in Fig.\ 1.  Median
magnitudes are plotted for stars detected in all 3 \mr625\ and all 3
\mb435\ images.  Sources within 0\secspt 5 and 1\secspt 0 of the
boresite-corrected X-ray source position are indicated as triangles
and squares, respectively.  The proposed optical counterpart to the
qLMXB is distinctly blue and \ha-bright (filled triangle).  Three
BY~Dra candidates are also detected (small circles).  Approximate
magnitude limits are indicated with dashed lines.  Several probable
white dwarfs are visible in the lower left of the left-hand panel.}
\vspace*{0.5cm}
\label{fig02}
\end{figurehere}

\section{Results}

Color--magnitude diagrams (CMDs) for the 1454 stars in the 20\asec\
\x\ 20\asec\ field are shown in Fig.\ 2.  To reduce the effects of any
remaining cosmic rays, we have plotted median magnitudes.  Stars are
shown if they appear in all 3 \mr625\ and all 3 \mb435\ images and any
number of \ha\ images.  This ensures that blue stars (e.g., white
dwarfs) are not overlooked, even if they are too faint to be detected
in \ha.

In the \br\ vs. \mr625\ diagram (left panel), the main sequence can be
seen extending $\simgreat$ 6 magnitudes below the turnoff and about a
dozen white dwarfs are visible at the lower left from \mr625\ $\simeq
23.5-25.5$ and \br\ $\simeq 0-1$.  In the right panel we plot \hr\
vs. \mr625.  Here the main sequence appears nearly vertical and has
been shifted so that it is approximately centered on \hr\ $=$ 0.  In
both diagrams, stars within 0\secspt 5 and 1\secspt 0 of the
boresite-corrected Chandra position are indicated with triangles and
squares, respectively.

The most interesting star in the X-ray error circle is shown as a
filled triangle in Fig.\ 2.  The star is very faint, at \mr625 $=$
25.2 and \br\ $=$ 1.5, and would not have been detected in \mb435\ or
\ha\ were it not considerably blue and \ha-bright.  It is $\simgreat$
1.5 magnitudes bluer than the main sequence and noticeably redder than
white dwarfs of comparable brightness.  In the \hr\ diagram, the star
lies \about 1.3 magnitudes to the left of the main sequence,
suggesting the presence of a strong emission line.  Despite being so
faint, the star was found in all 10 images by the automated
DAOPHOT/FIND routine, making it one of the most reliable detections
with \mr625 $>$ 25.  No other star in the 1\asec\ error circle is
significantly blue or \ha\ bright.  We identify this star as the
probable optical counterpart of the X-ray source.

A finding chart for the proposed optical counterpart is shown in Fig.\
3.  It lies just $\Delta$r$^\prime$ = 0\secspt 25 from the
boresite-corrected X-ray position (Table 1, column 5).  Even without any
boresite correction, the X-ray and optical positions are offset by
0\secspt 40 (Table 1, column 4)---well within the \about 1\asec\
uncertainty of HST and Chandra absolute positions.  That the agreement
between the optical and X-ray positions improves after the boresite
correction further supports the identification of this star as the
optical counterpart.

\begin{figurehere}
\vspace*{0.5cm}
\centerline{\resizebox{3in}{!}{\includegraphics{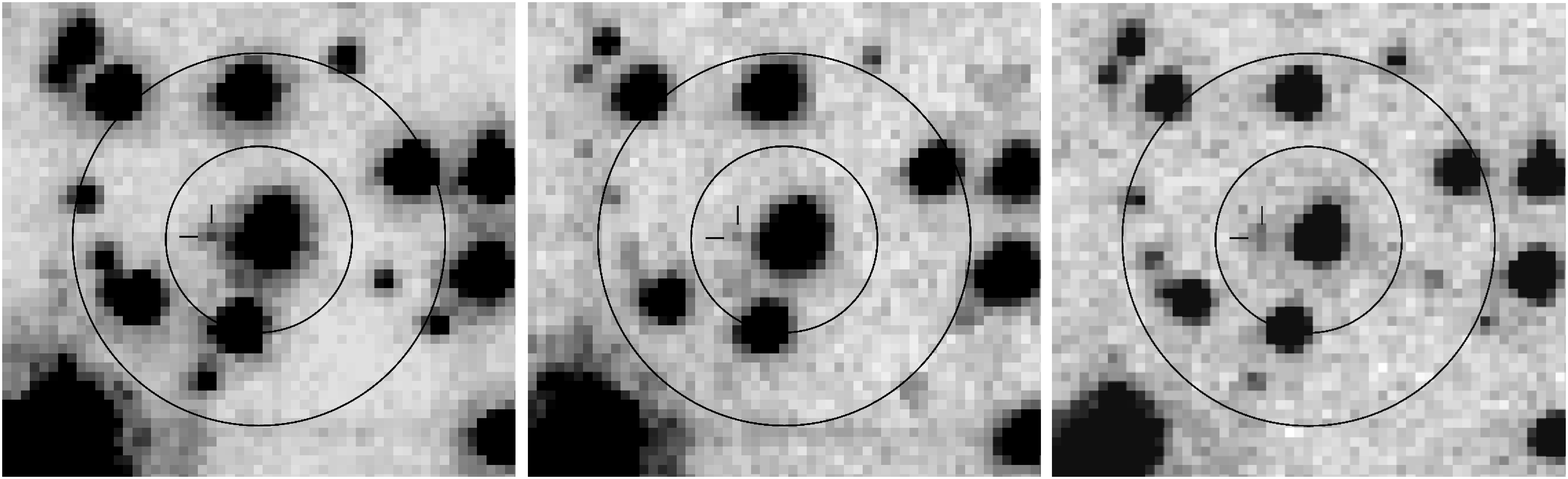}}}
\vspace*{0.2cm}
\caption{Finding chart for the proposed optical counterpart to the
qLMXB (see cross hairs) in \mr625\ (left), \mb435\ (middle), and \ha\
(right) filters.  These images are averages of the 3 or 4 available
exposures in each filter.  Error circles of radius 0\secspt 5 and
1\secspt 0 centered on the boresite-corrected position are shown.}
\vspace*{0.5cm}
\label{fig03}
\end{figurehere}

We performed several further tests to check on the reality of this
faint object and the reliability of the measured magnitudes.  First,
we examined the star visually in each of the ten long-exposure images,
and verified that no cosmic rays had affected the photometry.  In CMDs
made from averaged instead of medianed magnitudes, the star appears in
similar locations.  Comparable results were also obtained using
ALLFRAME, which is somewhat less susceptible to contamination of faint
star magnitudes by bright neighbors.  This is reassuring, since the
candidate has a neighbor just 0\secspt 3 (6 pixels) away that is
\about 5.5 magnitudes brighter in \mr625\ (Fig.\ 3).  We also examined
the residuals in the vicinity of the star after subtracting it out of
each of the images and found that the subtractions were clean,
with no sign of any flux being left behind.  Thus the object appears
to be stellar.

As a final test, we computed standard deviations for magnitudes
measured in each filter (Fig.\ 4).  These plots give an indication of
the accuracy of the photometry as a function of magnitude and also
show that, in all filters, the star is $\simgreat$ 0.5 magnitudes
brighter than the faintest stars detected.  The accuracy with which it
is measured ($\sigma$ $\simless$ 0.2 mag in all filters) implies that
unusual \br\ and \hr\ colors are clearly significant.  That the star
is blue and \ha-bright relative to other stars of comparable \mr625\
magnitude can also be verified visually (Fig.\ 3).

\begin{figurehere}
\vspace*{0.8cm}
\centerline{\resizebox{3in}{!}{\includegraphics{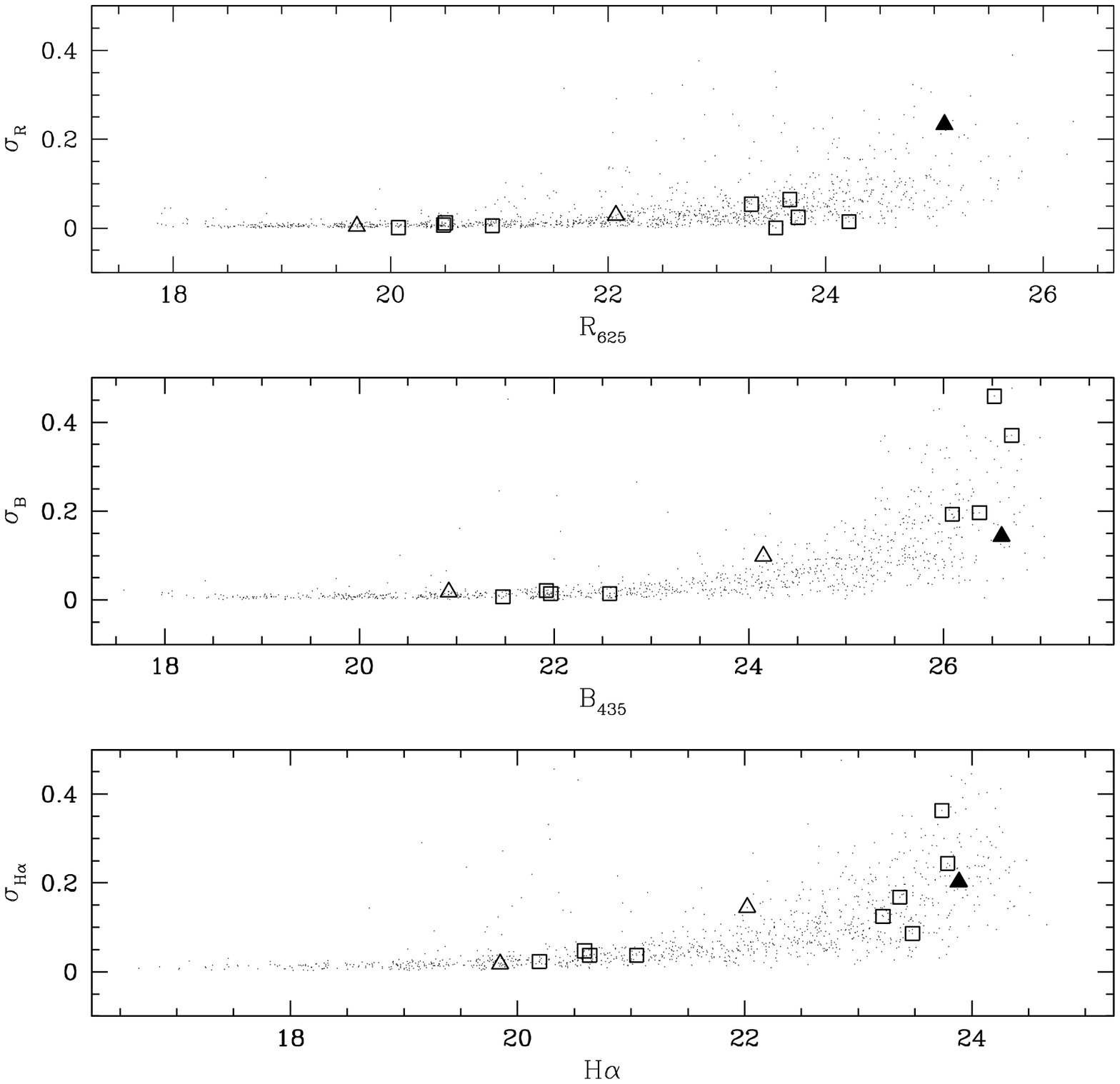}}}
\vspace*{0.2cm}
\caption{Plot of sigma vs.\ median magnitude for the \mr625, \mb435\
and \ha\ filters.  Detection in all of the ``long'' exposures in a
given filter was required for inclusion in the corresponding plot.
Symbols are as in Fig.\ 2.  Sigma values measured for the proposed
optical counterpart are consistent with being due to measurement
uncertainties alone.}
\vspace*{0.8cm}
\label{fig04}
\end{figurehere}

\section{Discussion}

The star we have identified as the optical counterpart of the qLMXB is
faint, blue and \ha-bright.  These characteristics are typical of
semi-detached binaries in which a compact object is accreting from a
low-mass companion (e.g., Cool et al.\ 1998).  From the observed 1.3
magnitude \ha\ excess and the widths of the F658N and F675W filters,
we infer the presence of an emission line with an equivalent width of
EW(\ha) $\simeq$ 220\AA, after correcting for the contribution of the
line to the flux through the \mr625\ filter.  Such a strong line is
reminiscent of lines seen in short-orbital-period cataclysmic
variables with low accretion rates (Patterson 1984, Tylenda 1981).
Comparably strong \ha\ emission is also seen in the SXT GRO J0422+32,
a black hole candidate with P$_{orb}$ \about 5 hrs (Harlaftis et al.\
1999).  Whereas in principle the neutron star might have been
accreting from the cluster interstellar medium (Rutledge et al.\
2002), the characteristics of the optical counterpart strongly suggest
the presence of an accretion disk and a low-mass binary companion.

The object's X-ray to optical flux ratio is high compared to qLMXBs in
the field but is on the order of those detected or constrained
optically in globular clusters.  Assuming a $V-R$ color of \about 0.5,
we estimate \fxfv\ \about 240 for this object.  Similar estimates for
the two qLMXBs in 47~Tuc yield \fxfv\ \about 50 for X5 and a lower
limit of \fxfv\ \simgt 180 for X7 (the latter based on a limit on the
brightness of the optical counterpart; Heinke et al.\ 2003, Edmonds
et al.\ 2002).  Our limit on the magnitude of the optical counterpart
of the qLMXB in NGC~6397 implies \fxfv\ \simgt 50 (Grindlay et al.\
2001).  Field qLMXBs Aql~X-1, Cen~X-4 and SAX~J1808 have \fxfv\ \about
$10-20$ judging from information given by Campana et al.\ (1998),
Chevalier et al.\ (1989, 1999), and Homer et al.\ (2001).

Further insight can be gained from the intrinsic color and absolute
magnitude of the optical counterpart.  Adopting a reddening of E(B-V)
= 0.11 (Lub 2002), we derive extinction values of A$_{435}$ = 0.45 and
A$_{625}$ = 0.29 and thus an intrinsic color of (\mb435 $-$
\mr625)$_0$ = 1.3.  This color helps to rule out the possibility
raised by Rutledge et al.\ that the source could be a narrow-line
Seyfert I galaxy.  For a plausible spectrum, it corresponds to a Sloan
$g^\prime - r^\prime$ color of \about 1.0, which is redder than 150
such galaxies studied by Williams et al.\ (2002).  The high \fxfv\
ratio further argues against its being an active galaxy, as AGN
typically have \fxfv\ values $1-2$ orders of magnitude lower (e.g.,
Keokemoer et al.\ 2004), as does the excess \ha\ emission, since a
strong line would have to be redshifted into the F658N bandpass by
chance.  A measurement of the object's proper motion would help
confirm its cluster membership, which would also rule out the
possibility that it is a field qLMXB along the line of sight to \wcen.
We estimate that a baseline of \simgt 5 years would be required to
make such a measurement given the faintness of this object and the
\about 6 mas/yr (\about 0.1 WFC pixels/yr) proper motion of \wcen.

As a cluster member, the star is at \about 5.0 kpc.  Using
the extinction above, the distance modulus is then (m$-$M)$_{625}$ =
13.8, yielding an absolute magnitude of \MR625\ = 11.4.  Adjusting for
the \ha\ emission line within the F625W bandpass gives \MR625 $\simeq$
11.6 for the continuum alone.  A $\pm$0.5 kpc uncertainty in the
distance (cf.\ van Leeuwen et al.\ 2000 vs.\ Thompson et al.\ 2001)
introduces an uncertainty of about $\pm$0.2 mag in these values.

This is the faintest optical counterpart detected for a qLMXB.  The
only known system that may be fainter is the one in NGC~6397, for
which a limit of $M_V$ $>$ 11 has been derived from its non-detection
in HST/WFPC2 images (Grindlay et al.\ 2001).  The optical counterpart
to X5 in 47~Tuc and to the field qLMXBs Aql~X-1 and Cen~X-4 ($M_V$ =
8.2, 8.1, and $7.5-8.5$, respectively; Edmonds et al.\ 2002 and
references therein) are all considerably brighter.  Such a faint
absolute magnitude implies either that the secondary star is of very
low mass or that it is a compact star.  In the case of a main-sequence
secondary, we can obtain an upper limit to the mass by assuming that
it provides all the light in the \mr625\ band.  For the metallicity of
\wcen, an absolute magnitude of \MR625 = 11.6 corresponds to a
zero-age main-sequence (ZAMS) star with a mass of \about 0.14\msun\
(Baraffe et al.\ 1997).  Such a star would contribute about 20\% of
the light in the \mb435\ band, leaving a disk with \MB435 $\simeq$
13.1.

We cannot place a lower limit on the mass of the secondary star.  However,
the intrinsic color of the system is sufficiently red, despite its
location blueward of the main sequence, that it appears likely that at
least some, if not all, of the light in the \mr625\ band comes from
the secondary.  To account for its color with a power-law spectrum
from a disk alone would require a positive exponent: f$_\lambda$
\about\ $\lambda ^{0.15}$.  A more typical disk spectrum going as
$\lambda ^{-2}$ would have an intrinsic color of \intbr\ \about 0.2
and account for only \about 40\% of the light in the \mr625\ band, even
if all the \mb435\ flux were from the disk.  The remaining \mr625-band
light could be accounted for by a main-sequence star of mass \about
0.12\msun.  The difficulty of accounting for the object's color
without a significant contribution from the secondary star argues
against the possibility that the secondary is compact.

A main-sequence secondary star mass of $\simless 0.12-0.14$\msun\
implies a short orbital period for the system.  In particular, for a
ZAMS star in this mass range to fill its Roche lobe requires an
orbital period of $\simless$ $1.2-1.5$ hr (Warner 1995); longer
periods are possible if the star is underdense (e.g., Edmonds et al.\
2002, Kaluzny \& Thompson 2003).  

The disk in the system must also be intrinsically very faint, with
\MB435 \about 13.  This is considerably fainter than even the disk in
SAX J$1808.4-3658$, a transient with a 2-hr period, whose disk is
estimated at M$_B$ = $5.7-9.0$ (Homer et al.\ 2001).  Disks this faint
are seen in CVs with periods below the period gap, i.e., $\simless$ 2
hrs (e.g., Sproats et al.\ 1996), with mass transfer rates of order of
10$^{-11}$ \msunyr\ driven by gravitational radiation (Warner 1995).
That an accreting neutron star, with a potential well a factor of
\about 1000 larger, would have such a faint disk suggests an extremely
low rate of accretion onto the neutron star at present, on the order
of 10$^{-14}$ \msunyr.  This is significantly lower than the rate
driven by gravitational radiation, if indeed the system has a short
orbital period.  A possible explanation for this apparent discrepancy
could be that material is being transferred to the disk from the
secondary, but that little, if any, is accreting onto the neutron star
from the disk at the present time.  Such a picture would fit with the
scenario proposed by Rutledge et al., in which the observed X-ray
emission arises from the neutron star itself, and not from present-day
accretion.  In the Brown et al.\ (1998) model, deep crustal heating of
the neutron star's core occurs during transient accretion events; a
time-averaged mass transfer rate of \about 8 \x\ 10$^{-12}$ \msunyr\
is required to explain the current X-ray luminosity of the neutron
star in this system.  We observe no variability in the optical flux
($\simless$ 0.2 mag, 1 $\sigma$---Fig.\ 4), which is also consistent
with little or no current accretion onto the neutron star.  However,
the constraints we place are limited by the faintness of the star, the
small number of exposures in each filter and the short (\about 1.5 hr)
time span of the observations.

Finally, we note that several other stars of interest appear in the
CMD in Fig.\ 2.  Three stars appear mildly \ha-bright in the right
panel but are on or near the main sequence in the left panel (small
circles).  Visual inspections suggest the stars are well measured.
These may be BY Dra-type active stars as have been seen, e.g., in
NGC~6397 (Taylor et al.\ 2001).  About a dozen faint blue stars with
\br $\simless$ 1 and \mr625 $\simgreat$ 23.5 are probable white dwarfs.
Finding such a large number in this small region suggests that of
order a thousand white dwarfs should be detected in the full ACS
mosaic.

We gratefully acknowledge discussions with Scott Anderson and thank
Marco Sirianni for providing ACS/WFC calibration information in
advance of publication.  This work was supported by NASA grant
HST-GO-9442.

\end{document}